# ON THE QUANTUM-CLASSICAL CHARACTER OF THE QUANTUM WAVEFUNCTION OF MATERIAL PARTICLES

D. Dragoman – Univ. Bucharest, Physics Dept., P.O. Box MG-11, 077125 Bucharest, Romania, e-mail: danieladragoman@yahoo.com

ABSTRACT

We show that the quantum wavefunction, interpreted as the probability density of finding a single non-localized quantum particle, which evolves according to classical laws of motion, is an intermediate description of a material quantum particle between the quantum and classical realms. Accordingly, classical and quantum mechanics should not be treated separately, a unified description in terms of the Wigner distribution function being possible. Although defined on classical phase space coordinates, the Wigner distribution function accommodates the nonlocalization property of quantum systems, and leads to both the Schrödinger equation for the quantum wavefunction and to the definition of position and momentum operators.

INTRODUCTION

Quantum mechanics of material particles, i.e. particles with positive mass, which we refer to simply as particles in the following, differs qualitatively from classical mechanics in that it systematically replaces the fundamental concepts of the latter with completely different notions. For example, the state of a classical localized particle as determined by position and momentum vectors $\mathbf{r}$ and $\mathbf{p}$, respectively, is described in quantum mechanics by a vector $|\Psi\rangle$ in the Hilbert space, which in the Schrödinger formalism becomes a non-localized function $\Psi$ of position or momentum. In turn, the classical $\mathbf{r}$ and $\mathbf{p}$ vectors, with Cartesian components $x_i$ and $p_i$, respectively, become in quantum mechanics the operators $\hat{\mathbf{r}}$ and $\hat{\mathbf{p}}$, with Cartesian components $\hat{x}_i$ and $\hat{p}_i$, respectively. These quantum operators, which act on the quantum state (wavefunction), no longer commute but satisfy the commutation relation

$$[\hat{x}_i, \hat{p}_j] = \hat{x}_i \hat{p}_j - \hat{p}_j \hat{x}_i = i\hbar \delta_{ij}. \qquad (1)$$

The meaning of the wavefunction $\Psi$ is not yet clearly established, but it is generally considered that it is an appropriate tool for calculating the probability of outcomes of a measurement of an arbitrary quantum observable $A$, represented by an operator $\hat{A}$ with eigenfunctions $\psi_n$ and eigenvalues $A_n$. These eigenvalues are the only possible results of measurements of the observable $A$, the probability of obtaining them for a quantum system with a wavefunction $\Psi = \sum_n a_n \psi_n$ being proportional to $|a_n|^2$.

The aim of this paper is to show that the wavefunction in quantum mechanics is in fact a description of the quantum system that is intermediate between the quantum and the classical world. More precisely, the quantum wavefunction contains classical parameters in both the equation of motion and the corresponding initial conditions. Having established that the wavefunction is in fact a probability density of finding classically-moving non-localized particles, we then argue that the net separation between the classical and quantum mechanical formalism has no sense, and therefore the only appropriate description of quantum mechanics is the phase space formalism. The Wigner distribution function, although defined on classical phase space coordinates, provides a suitable link between the quantum and classical aspects of the wavefunction and allows the derivation of the Schrödinger equation for the wavefunction as well as of the form of the position and momentum operators acting on it.

QUANTUM ASPECTS THE QUANTUM WAVEFUNCTION

The wavefunction of a quantum particle in the position representation, $\Psi(\boldsymbol{r})$, satisfies the Schrödinger equation

$$i\hbar \frac{\partial \Psi}{\partial t} = \hat{H}\Psi = -\frac{\hbar^2}{2m}\nabla^2\Psi + V(\boldsymbol{r},t)\Psi \qquad (2)$$

where $t$ is the time coordinate, $m$ is the mass of the quantum particle, and the Hamiltonian operator $\hat{H}$ is obtained by replacing the position and momentum coordinates in the classical non-relativistic Hamiltonian $H(\boldsymbol{r},\boldsymbol{p},t) = \boldsymbol{p}^2/2m + V(\boldsymbol{r},t)$ with the operators $\boldsymbol{r}$ and $-i\hbar\nabla$, respectively, which act on $\Psi(\boldsymbol{r})$. Besides the wavefunction itself, which is commonly considered as quantum in character, there are two definitely non-classical quantities in (2): the parameter $i\hbar$ on the left-hand-side and the form of the momentum operator, $\boldsymbol{p} = -i\hbar\nabla$. Indeed, in contrast to classical mechanics, where $\boldsymbol{p} = md\boldsymbol{r}/dt$ and, in particular, for a free particle starting from the origin of the system of coordinates $\boldsymbol{p}$ is parallel to $\boldsymbol{r}$, in quantum mechanics $\boldsymbol{p}$ is normal to the surface $\Psi(\boldsymbol{r}) = $ const., in a similar manner as the wavevector in optics is normal to the wavefront. Moreover, a complete determination of the state of a classical particle at the moment $t$ requires two measurements at different times for the determination of the temporal derivative of *r* that appears in the definition of *p*, whereas the quantum wavefunction established at a given time moment characterizes completely the quantum state, the momentum operator being defined, however, as the gradient of the spatially dependent wavefunction.

The analogy of the momentum operator definition in quantum mechanics and wave optics (in the latter case an operator $\hat{\boldsymbol{p}} = -i(\lambda/2\pi)\nabla$ can be introduced [1] that is canonically conjugate in the Hamiltonian sense to *r*, so that the wavelength $\lambda$ is the analog of $h = 2\pi\hbar$ while the time coordinate in quantum mechanics is to be replaced in wave optics with a spatial coordinate) reminds us of the well-known argument that classical mechanics is an approximation to quantum mechanics analogously to the way geometrical optics is an approximation to wave optics. However, nobody has ever claimed that wave optics is of a fundamentally different nature than ray optics, and nobody has ever tried to separate the two mathematical theories of optics, as is the case with the classical and quantum mechanics. The reason is, perhaps, that the wave and ray theories have appeared at rather the same epoch and that the optical wavelength is measurable with classical means, whereas Planck's constant is not observable in every-day experiences and hence had no place in classical mechanics.

Not only does classical particle acquire a wave-like behavior in quantum mechanics, but its phase space topology changes; for simplicity we restrict ourselves in the following to the one-dimensional case, in which the position is denoted by *x* and momentum by *p* ($\hat{p} = -i\hbar\partial/\partial x$ in quantum mechanics). Then, in the two-dimensional (*x*, *p*) phase space a classical particle is represented by a point, with an associated outer product $x \wedge p = (xp - px)/2 = 0$ since *x* and *p* are numbers, while in quantum mechanics, according to the commutation relation, $\hat{x} \wedge \hat{p} = i\hbar/2$. Since the outer product represents the oriented area of the parallelogram with sides $\hat{x}$ and $\hat{p}$ (see, for example [2]), it follows that a quantum particle is always localized in a phase space area $\hbar/2$; it is no longer a point, as in classical mechanics. (A similar change in the phase space topology occurs in wave optics with regard to ray optics.) The existence of a minimum phase space area bestows to the quantum wavefunction its wave-like behavior and, in particular, implies the possibility of interference of material particles, a property that distinguishes a quantum particle from a classical one.

This change in phase space topology, determined by the commutation relation (1), which is independent of the wavefunction, is the essence of quantum behavior. The specific form of the wavefunction intervenes only in the uncertainty relation,

$$\Delta x \cdot \Delta p \geq \hbar/2 \qquad (3)$$

where the uncertainty in $x$ is defined by $(\Delta x)^2 = \langle (x - \langle x \rangle)^2 \rangle = \langle x^2 \rangle - \langle x \rangle^2$, a similar relation existing for $\Delta p$, with the expectation value of any function of $x$ and $p$, $f(x, -i\hbar \partial/\partial x)$, calculated as $\langle f(x, -i\hbar \partial/\partial x) \rangle = \int \Psi^*(x,t) f(x, -i\hbar \partial/\partial x) \Psi(x,t) dx$; of course, when products of $x$ and $p$ are encountered, their order is not arbitrary. The uncertainty in $x$ ($p$) can be regarded as the spatial extent of the wavefunction in position (momentum). Note that both the commutation and uncertainty relations are defined in phase space.

CLASSICAL ASPECTS OF THE QUANTUM WAVEFUNCTION

In the Schrödinger equation (2), as in any formulation of quantum mechanics, the time coordinate $t$ and the mass $m$ of the quantum particle are purely classical parameters, with no associated operators. Moreover, the argument $r$ of the wavefunction and of the potential energy in (2) is no different from the classical position vector. An insightful demonstration that, for a quantum system interacting with an environment that can be characterized by semiclassical dynamical variables, the time coordinate in (2) originates in the classical evolution of the environment variables and the wavefunction of the quantum system depends on the environmental position variables (although this dependence is not explicit in the Schrödinger equation), can be found in [3].

If the classicality of time, mass and position in (2) is common knowledge, the identification of classical aspects of the wavefunction itself is not. We have established in the previous section that the form of the wavefunction is irrelevant for the commutation relation (1). But, what determines the form of the wavefunction? It is the preparation process, which involves the particle source and other filtering devices, that is responsible for the form of the initial quantum state $|\Psi_{in}\rangle$. Then, propagating through the set-up the wavefunction interacts continuously with different parts of the set-up, for example filters and slits, and retains information about them, as discussed in [4,5], the filtering devices having often a contribution to the form of $\Psi_{in}$. For example, in interference experiments involving slits the preparation device consists of the particle source <u>and</u> the slits, which are considered as a position-measuring device [6]. After passing through a slit the wavefunction is projected into a (classical) position state, (the position being that of the slit) which is considered as the initial wavefunction $\Psi_{in}$. Analogously, multiple slits generate a superposition of position states, the initial wavefunction (and hence the fringe pattern) depending on the (classical) distance between the slits. Detectors placed after the slit plane, and which make a determined (classical) angle $\theta$ with a certain slit, measure the momentum eigenstates [6], determining the final state of the wavefunction, $|\Psi_{fin}\rangle$. Quantum mechanics provides the probability $P=|\langle\Psi_{fin}|\Psi_{in}\rangle|^2$ that the initial state is found in the observed final state. Quantum interference is an attribute of this probability function and not of quantum particles.

In general, the effect of the components of the set-up (particle sources, filters, slits, detectors, etc.) on the wavefunction is treated qualitatively, though in many cases in agreement with experiments. Often at least the detecting device is downgraded to a semi-classical or even a classical device. Moreover, in practically all cases, the engineering of the quantum state of a particle implies interaction with classical fields or, at least, the fact that classical parameters intervene in the form of $\Psi_{in}$. For example, preparation of the motional ground state of a trapped ion and the engineering of its quantum state of motion is performed by applying electromagnetic fields, which drive transitions between ion's states, the occupation of the metastable excited state depending on classical parameters such as the duration of the excitation pulses and the wavelength detuning of the exciting laser from the ionic transition frequency [7] (see also the references in [8] for other methods of engineering the states of trapped ions). Another, the so-called projection synthesis method, can be used to sculpt an arbitrary motional state of a trapped ion from a previously prepared coherent motional state using appropriately chosen laser pulses, the durations and phases of the laser pulses determining the fidelity of the process [8]. Deterministic entanglement of trapped ions (entanglement of all ions at a specific time) can be engineered by applying laser beams with appropriately detuned frequencies and specific time durations, as well as electric fields on Paul traps [9] (see also the references in [9] for other entanglement experiments). The transition frequencies between quantum states of Rydberg atoms are tuned using electric fields, and superposition of energy states of such atoms are realized applying pulses of classical microwave radiation, atom-cavity and atom-atom entangled states being created and manipulated by combining these classical pulses with cavity fields with precisely determined frequency [10]. Moreover, atoms can be trapped, guided, and focused by the forces exerted by intense and inhomogeneous electromagnetic fields with appropriately chosen wavelengths or by current carrying wires, microfabricated dispersive elements such as Fresnel zone plates are used in atom and electron optics experiments, electrons are manipulated by classical electric and magnetic fields, in temporal interference experiments involving massive particles the wavefunction is determined by the classical delay between particle wavepackets, which also determines the parameters of temporal lenses and mirrors for atoms, and so on (see the extensive reference list in [11]).

In conclusion, the macroscopic environment influences the wavefunction from its generation up to its detection; the macroscopic world enters the quantum mechanical description through the parameters in the Schrödinger equation (time, position variables, often potential energy expressions) and through boundary conditions (the form of the initial wavefunction), the eigenstates of the observable to be measured (determined by a generally classical measuring apparatus decided upon by the experimenter) determining the outcome probabilities of possible measurement results, which test the validity of quantum mechanics. It is hazardous, therefore, to consider the Schrödinger equation as a purely quantum description of the system. A more appropriate interpretation of the Schrödinger equation as a description of a quantum particle, intermediate between the microscopic (quantum) and macroscopic (classical) realms, should be considered instead.

Another argument for the existence of a classical aspect of the wavefunction is based on the misleading assertion that the addition law of classical probabilities is invalidated by quantum interference phenomena. But is this really the case? In Ref. 12 it is convincingly demonstrated that the well-known fact in a two-slit experiment, namely that the probability of the photon reaching a certain point on the screen when both slits are open is not equal to the sum of probabilities of photons reaching that point when only one of the slits is open does not contradict the classical addition law, because what must be added are the probabilities of the photon passing through each one of the slit when <u>both</u> are open (similar generalizations hold for $n$-slit interferometers). And when this addition is performed classically we obtain the so-called quantum result, i.e. interference fringes on the screen.

Moreover, the quantum mechanical postulates are intrinsically related to classical measuring devices. More precisely, the quantum mechanical postulate, which states that a measurement of an observable *A* described by an operator $\hat{A}$ always yields one of the eigenvalues of $\hat{A}$, is not fulfilled in general. Its validity (and hence the conceptual foundations of quantum mechanics) depends on the strength of the interaction between the quantum system and the measuring apparatus, and, even worse, in weak measurements the result can even lie outside the range determined by the operators' eigenvalues [13,14]. The so-called strong measurements, which fulfill the quantum mechanical postulate, are those in which the measuring apparatus behaves classically, i.e. is prepared in a classical limit state for which the position uncertainty of the apparatus and thus the spatial extension of its wavefunction must be much larger than its de Broglie wavelength. Then, the real part of the Schrödinger equation satisfied by the wavefunction of the apparatus reduces to the Hamilton-Jacobi equation and vice-versa.

MEANING OF THE QUANTUM WAVEFUNCTION

The commutation relation together with (3) establish that the product of the uncertainties in *x* and *p*, and hence the extent of the wavefunction in phase space, must be equal to or larger than the minimum phase space area $\hbar/2$. Actually, the state representing a quantum particle cannot be localized in a phase space region smaller than a quantum blob, which is canonical invariant [4,15]. In our opinion the wavefunction (in the position representation) is the spatial probability density of finding a single quantum particle, which evolves according to classical laws but differs from a classical particle because it cannot be localized in a phase space area smaller than $\hbar/2$ (smaller than $(\hbar/2)^n$ for an *n*-dimensional wavefunction). In this sense, contrary to the position defended by Schrödinger [16] and by de Broglie, who postulated the coexistence of phase-matched particles and guided-waves [17], the wavefunction has no ontological meaning; it does not exist as a real entity independent of our knowledge about it, as do, in our opinion, quantum particles. The wavefunction has, however, an epistemological character (as argued by Born [18]) since it can be used to predict the results of future experiments from the knowledge of the present state of the system and the configuration of the measuring set-up through which it will pass.

The dissociation in meaning of the wavefunction from the quantum particle is imposed by experiments, which show that the wavefunction has a spatial extension that is not compatible with the localized character of the detected quantum particle. For example, in quantum interference experiments, the wavefunction must extend over the slits in order to observe interference fringes, but quantum particles pass through the slits. The wavefunction cannot represent the quantum particle, whose character does not change even if the wavefunction is modified by canonical transformations, for example. The existence of the minimum phase space area $\hbar/2$ does not necessarily mean that the quantum particle has this extent; it simply implies that a quantum particle cannot exist unless this phase space area is available. The minimum phase space area should be seen as a sort of self-sustained correlation area, necessary for the stability of the quantum particle. The extent of the quantum particle itself is not a question that present-day quantum mechanics can answer.

The wavefunction interpretation as a probability amplitude for a single quantum system as opposed to a parameter that acquires physical meaning only as ensemble average over a large number of identical systems at a given moment in time was already defended in [19]. In this reference the wavefunction $\Psi$ of a single particle acquires realistic interpretation through a so-called protective (or weak) measurement, i.e. a long-time measurement during which the wavefunction does not collapse because it interacts in a suitable manner with the environment, and during which the expectation values of several, even non-commuting observables can hence be determined. The wavefunction of the quantum system can eventually be recovered (up to a phase factor) from measurements of a sufficient number of observables, although all measurements are performed on a single system, and not on an ensemble of identical copies of the quantum system existing in various allowed states. The interaction term $H_{\text{int}}$ in the total Hamiltonian $H = H_0 + H_a + H_{\text{int}}$, where $H_0$ is the unperturbed Hamiltonian of the quantum system and $H_a$ is the Hamiltonian of the apparatus, depends generally on the coordinate of the apparatus, the possible states of which after interaction are entangled with the eigenstates of the measured observable. An actual observation of the state of the apparatus implies a collapse of the wavefunction into one of these eigenstates. No such collapse is possible in a protective measurement since the entanglement is absent. Protective measurements can take place when the state $|\Psi(t)\rangle$ of the quantum system remains almost unchanged during measurement. If $|\Psi(t)\rangle$ is at $t = 0$ and remains after the interaction time $T$ in a nondegenerate eigenstate of the total Hamiltonian $H$, the interaction is sufficiently weak such that $|\Psi(t)\rangle$ changes adiabatically and remains almost identical after $T \gg \hbar/\Delta E$, where $\Delta E$ is the smallest energy difference between $|\Psi\rangle$ and the other eigenstates of $H$. If $|\Psi(0)\rangle$ is not an eigenstate of $H$ it can be protected through repeated and dense measurements in the $[0,T]$ interval of an observable for which the evolution of $|\Psi(0)\rangle$ under the unperturbed Hamiltonian $H_0$ is a nondegenerate eigenstate. Protective measurements allow the distinction between protected nonorthogonal states of the quantum system, and can be generalized to determine the density matrix of a single quantum system [20].

    A further insight into the meaning of wavefunction, as corresponding to classically moving particles, is provided in [21]. In this reference it is shown that the wavefunction is the probability density of classically propagating particles starting from the observation that if geometrical optics exactly accounts for the propagation of field discontinuities, then classical mechanics should describe exactly the propagation of discontinuities in the quantum wave function. And, indeed, the eikonal equation in optics is found to correspond to the classical Hamilton-Jacobi equation, the discontinuities propagating along the rays in geometrical optics

and along classical trajectories in relativistic quantum mechanics (the nonrelativistic case can also be accommodated) at least in the non-dispersive, time-harmonic wave case [21].

A further support for the classical propagation law of quantum particles is provided in [22], where the evolution laws of quantum observables are derived from the classical Hamilton equations with the only additional assumption of phase space non-commutativity, which implies that (1) is satisfied by $x$ and $p$ and that the quantum particle occupies a phase space area of $\hbar/2$. This result is obtained without the introduction of Hilbert spaces and quantum operators and is independent of the quantum mechanical formalism. It states that the quantum particle differs from the classical one only with regard to phase space localizability.

Because the quantum particle, understood throughout this paper as a concentration of energy, is not classical (in the sense that it is not point-like), although satisfying classical laws of motion, our interpretation of the wavefunction differs from that in the hidden-variable Bohmian quantum mechanics, which implies classical particles that follow trajectories perfectly determined by the classical equation of motion complemented with an additional wavefunction-dependent quantum potential [23].

Another argument toward our interpretation of the wavefunction is that the classical limit of a quantum state is an ensemble of orbits and not just a single orbit [24], as is usually assumed by a superficial interpretation of the Ehrenfest's theorem. The classical limit of quantum states implies then that quantum averages agree with classical averages and probability distributions in quantum mechanics concur with classical probabilities. In its standard formulation Ehrenfest's theorem states that the expectation values of the position and momentum operators follow a classical trajectory, i.e. $d\langle\hat{q}\rangle/dt = \langle\hat{p}\rangle/m$, $d\langle\hat{p}\rangle/dt = \langle F(\hat{q})\rangle \cong F\langle(\hat{q})\rangle$, where $F = -\nabla V$ is the force associated to the scalar potential $V$ in the Hamiltonian operator $\hat{H} = \hat{p}^2/2m + V(\hat{q})$.

RECONCILIATION OF THE CLASSICAL AND QUANTUM ASPECTS OF THE QUANTUM WAVEFUNCTION

The mainstream quantum mechanics states that the quantum and classical descriptions of the world are completely different, and at best correspondences must be searched between them. We have seen in the previous sections, however, that the quantum wavefunction carries information about the classical environment, and therefore the classical world is in (permanent) contact with the quantum world. How can we reconcile these two points of view?

A straightforward answer is provided by the Wigner distribution function (WDF) formalism of quantum mechanics, which employs classical phase space coordinates. More precisely, for pure quantum states the WDF is defined as [25]

$$W(x,p;t) = h^{-1} \int \Psi^*\left(x - \frac{x'}{2}; t\right) \Psi\left(x + \frac{x'}{2}; t\right) \exp(-ipx'/\hbar) dx', \qquad (4)$$

(a similar definition exists for mixed states, but we do not refer to it here since we are interested only in the wavefunction) and provides the same information about the quantum state as the wavefunction, the phase space formalism being analogous to that of Heisenberg and Schrödinger. Among the many useful properties of the WDF, reviewed in [26-30], we mention the fact that the wavefunction can be recovered (up to a constant factor) from the WDF through an inverse Fourier transform: $\Psi(x)\Psi^*(0) = \int W(x/2, p) \exp(ipx/\hbar) dp$ and that the spatial probability density in quantum mechanics is given by: $|\Psi(x)|^2 = \int W(x, p) dp$. The fact that the WDF, although real, is not positive defined, as a distribution function in classical mechanics, should not be a problem in quantum mechanics (as well as wave theory), since negative regions of WDF, corresponding in wave optics to tamasic or "dark" rays, are necessary to accommodate phenomena such as interference and diffraction [31]. An alternative set of postulates to that of standard quantum mechanics has been recently defined in terms of the WDF [4], including an interpretation of the measurement problem that allows the inclusion of classical measuring devices in the phase space formalism of quantum mechanics. So, the classical and quantum aspects of the physical world are best reconciled in phase space, which use the same classical coordinates for both descriptions. The phase space representation of quantum mechanics provides even a refined quantum-classical correspondence rule than the Bohr-Heisenberg correspondence principle, as shown in [32]. More precisely, using the WDF it is not only possible to demonstrate that the frequency and probability of a quantum mechanical transition between stationary states correspond, respectively, to a multiple of the classical dynamic frequency and to a classical Fourier coefficient, but (even neglecting higher-order terms in the evolution equation for WDF) the WDF associated to the transition is limited to a phase-space trajectory with an average energy between the energies of the stationary states.

As stated in the previous sections, the essence of quantum mechanics is captured by the commutation relation (1). In the phase space, however, $x$ and $p$ commute, since they are numbers. Nevertheless, the commutation relation is incorporated in the very definition of the WDF, as we show below by deriving both the Schrödinger equation and the uncertainty relations from the evolution law of the WDF. To our knowledge, this derivation is performed here for the first time.

The evolution law of the WDF

$$\frac{\partial W}{\partial t} + \frac{p}{m}\frac{\partial W}{\partial x} - \frac{\partial V}{\partial x}\frac{\partial W}{\partial p} = \sum_{n=1}^{\infty} \frac{(\hbar/2i)^{2n}}{(2n+1)!} \frac{\partial^{2n+1}V}{\partial x^{2n+1}} \frac{\partial^{2n+1}W}{\partial p^{2n+1}} \tag{5}$$

is usually derived from the Schrödinger equation (2). However, the process can be reversed, and (2) can be derived from (5), since it can be demonstrated that the Fourier transform of the real-valued WDF over the momentum variable can always be written as [33]

$$Z(x, x'/2; t) = \int W(x, p; t) \exp(ipx'/\hbar) dp = \Psi^*(x - x'/2; t)\Psi(x + x'/2; t). \tag{6}$$

(The demonstration in [33] refer to a classical, positive-definite WDF that satisfies the Liouville equation, i.e. the left-hand-side of (5), but only the real-valuedness of the WDF is necessary to show (6).) In order to derive from (5) the Schrödinger equation, we follow the treatment in [34], but starting from completely different premises: in our case the system is completely described by the quantum WDF (4) instead of a positive-defined phase space probability density for point-like classical particles, which follows the evolution law (5) (the classical WDF follows the Liouville equation). First, we derive the evolution law for the Wigner-Moyal transformation $Z(x, x'/2; t)$ in (6), which can be considered as a characteristic function over the momentum space. This equation is

$$\frac{\partial Z}{\partial t} - i\frac{\hbar}{m}\frac{\partial^2 Z}{\partial x \partial x'} + i\frac{x'}{\hbar}\frac{\partial V}{\partial x}Z = -\sum_{n=1}^{\infty} i\frac{x'}{\hbar}\frac{(x'/2)^{2n}}{(2n+1)!}\frac{\partial^{2n+1}V}{\partial x^{2n+1}}Z, \tag{7}$$

which can be expressed in the new variables $y = x + x'/2$, $y' = x - x'/2$ as

$$\left[\frac{\hbar^2}{2m}\left(\frac{\partial^2}{\partial y^2} - \frac{\partial^2}{\partial y'^2}\right) - [V(y) - V(y')]\right]Z(y, y'; t) = -i\hbar\frac{\partial}{\partial t}Z(y, y'; t), \tag{8}$$

which is the analog of the Schrödinger equation for the quantum wavefunction. The latter is then finally obtained from (8) and (6).

A few remarks are in order: the derivation of the Schrödinger equation from the characteristic function $Z$ was performed in [34] for the particular case in which $x'$ is an infinitesimal parameter. This condition and the fact that the starting equation was the Liouville equation for the phase space probability density instead of (5) lead to the neglect of the right-hand-side of (7) in [34]. If $x'$ would have been negligible, we could in turn have neglected the right-hand-side of (7), the resulting equation appearing as derived from the Liouville equation.

As mentioned above, for the classical phase space coordinates $x$ and $p$ the commutator vanishes, i.e. $[x, p] = 0$, as applied on the WDF, which means that the WDF is well defined on any phase space point. However, the uncertainties in $x$ and $p$ still satisfy (3) if the expectation value (ensemble average) is understood as phase space average. More exactly, defining the phase space expectation value of a function of $x$ and $p$ as $\overline{f(x,p)} = \int f(x,p)W(x,p)dxdp$ (the WDF is considered normalized) the average phase space values for the position and momentum operators become identical to the expectation values for the operators $\hat{x} = x$ and $\hat{p} = -i\hbar\partial/\partial x$ in standard quantum mechanics, respectively, acting on the quantum wavefunction:

$$\overline{x} = \int \Psi^*\left(x - \frac{x'}{2};t\right) x \Psi\left(x + \frac{x'}{2};t\right) \exp(-ipx'/\hbar) dx'dxdp = \int \Psi^*(x;t) x \Psi(x;t) dx = \langle \hat{x} \rangle \quad (9a)$$

$$\overline{p} = \int \Psi^*\left(x - \frac{x'}{2};t\right) p \Psi\left(x + \frac{x'}{2};t\right) \exp(-ipx'/\hbar) dx'dxdp = \int \Psi^*(x;t)\left(-i\hbar\frac{\partial}{\partial x}\right)\Psi(x;t) dx = \langle \hat{p} \rangle$$
(9b)

In general, the expectation value of any function of $\hat{x}$ and $\hat{p}$ can be calculated as a phase space average with the WDF as a weighting function, i.e. as $\langle f(\hat{x},\hat{p}) \rangle = \overline{f(x,p)}$ $= \int f(x,p)W(x,p)dxdp$.

Equations (9a) and (9b) can serve to define the non-commutative position and momentum operators that act on the quantum wavefunction, $\hat{x}$ and $\hat{p}$, respectively, for which $[\hat{x},\hat{p}] = i\hbar$. The commutation relation (1) is replaced in the phase space treatment of quantum mechanics by commuting operators that act on an extended two-dimensional distribution function, which cannot be localized in a phase space area smaller than $\hbar/2$; the quantum parameter $\hbar$ is incorporated in the very definition of the WDF through the exponential term in (4), which relates the classical $x$ and $p$ variables through a Fourier transform describing a wave-like object extended in the $(x,p)$ plane. In summary, the accommodation of the quantum parameter $\hbar$ does not necessarily require the change of classical vectors into quantum mechanical operators; the replacement of classical point-like particles with extended quantum mechanical concentrations of energy suffice for this purpose.

The WDF thus reconciles the quantum and classical treatments of phenomena, and is perfectly compatible to both the Schrödinger equation and the commutation relation, although defined on the classical phase space. The WDF, as advocated in [4], can be used to treat in a unified manner both quantum and classical systems. The use of the WDF to describe quantum systems is also supported by experimental reasons. Marginal distributions of the WDF for homodyne observables are easily determined experimentally, and offer an equivalent description of the quantum state to that of the quantum wavefunction or density matrix. The (positive) marginal distribution of a measurable position observable $X$, which is a combination of quadratures $\hat{X} = \mu\hat{x} + \nu\hat{p}$, where the $\mu$ and $\nu$ parameters refer to different scaled and rotated reference frames in phase space in which $X$ is determined, is related to the WDF (for $\hbar = 1$) by

$$w(X,\mu,\nu) = (2\pi)^{-2} \int \exp[-ik(X - \mu x - \nu p)] W(x,p) dk dx dp, \tag{10}$$

or $W(x,p) = (2\pi)^{-1} \int \exp[i(X - \mu x - \nu p)] w(X,\mu,\nu) d\mu d\nu dX$. The WDF and the marginal distributions, which represent probability distributions for the position $X$ in the scaled and rotated phase space reference frames, determine each other completely; in consequence the marginal distributions offer equivalent information to the Schrödinger equation on the evolution of the quantum system and on the characteristics of its stationary states. Their evolution under a Hamiltonian $H = \hat{p}^2/2 + V(\hat{x})$ is described by

$$\frac{dw}{dt} - \mu\frac{\partial w}{\partial v} - 2\operatorname{Im} V\left[i\frac{v}{2}\frac{\partial}{\partial x} - \left(\frac{\partial}{\partial X}\right)^{-1}\frac{\partial}{\partial \mu}\right]w = 0 \qquad (11)$$

(see [35] for details). Marginal distributions with $\mu^2 + v^2 = 1$ are actually measured in tomography experiments.

CONCLUSIONS

We have argued that the quantum wavefunction is an intermediate description of a material quantum particle between the quantum and classical realms. The quantum wavefunction is in our view a probability density of finding a single quantum particle, which evolves according to classical mechanical laws but is not classical since it cannot be localized in a phase space area smaller than $\hbar/2$ (for a one-dimensional system). According to these considerations we feel that classical and quantum mechanics should be treated separately, a unified description being possible in terms of the WDF. This distribution function, although defined on classical phase space coordinates, includes in its definition the nonlocalization property of quantum systems, and leads to both the Schrödinger equation for the quantum wavefunction and to the definition of position and momentum operators acting on it.